\documentclass[showpacs,amsmath,amssymb, prb,twocolumn]{revtex4}
\usepackage{graphicx}
\usepackage{dcolumn}
\usepackage{bm}

\begin{document}

\title{Generalization of Blonder-Tinkham-Klapwijk theory to particle-hole mixing boundary conditions: $\pi$-shift and conductance dips}
\author{M. Catapano$^1$}
\author{F. Romeo$^1$}
\author{R. Citro$^{1,2}$}
\author{F. Giubileo$^2$}
 \affiliation{
  $^1$Dipartimento di Fisica "E. R. Caianiello", Universit\`a degli Studi di Salerno, I-84084
Fisciano (SA), Italy\\
  $^2$CNR-SPIN Salerno, Via Giovanni Paolo II, I-84084 Fisciano (SA), Italy}

\date{\today}

\begin{abstract}
We generalize the Blonder-Tinkham-Klapwijk theory considering non-diagonal boundary conditions in the Bogoliubov-de Gennes scattering problem, to describe anomalous conductance features often reported for normal-metal/superconductor contacts. We calculate the differential conductance spectra showing that conductance dips, not expected in the standard formulation, are explained in terms of phase $\pi$-shift, between the bulk and the interface order parameter, possibly induced by a localized magnetic moment.
A discretized model is used to give quantitative evaluation of the physical conditions, namely the polarization and transparency of the interface, needed to realize the phase gradient.
\end{abstract}

\maketitle

\section{Introduction}
Since the introduction of the point contact spectroscopy technique\cite{Naidyuk} in 70's, in which a micro-constriction is created pressing a metallic tip onto a superconducting sample, the study of normal-metal/superconductor (N/S) junctions has represented an important means for the comprehension of several physical phenomena at the interface. The BTK theory\cite{BTK}, formulated by Blonder-Tinkham-Klapwijk few years after, furnished a powerful tool to describe N/S contacts with transparency ranging from metallic to tunneling regime, the interface barrier strength being modeled using a Dirac delta potential of arbitrary amplitude. The theory, formulated in terms of Bogoliubov-de Gennes (BdG) equations\cite{deGennes}, provides the transmission and reflection coefficients and it succeeds in explaining the conversion of a quasi-particle current into a supercurrent, due to the Andreev reflection\cite{Andreev}, allowing accurate prediction of the experimental results about differential conductance spectra, energy gap and excess current.
More recently, some modified BTK models have been proposed in order to take into account spin polarization\cite{Soulen,Upadhyay,Strijkers}, diffusive contacts\cite{Mazin,Woods}, superconducting proximity effect\cite{Strijkers} and thin ferromagnetic layers at the interface\cite{Romeo1,Romeo2}. These formulations have been motivated by several experimental evidences\cite{Soulen,Srikanth,Zhuang,Xiong,Sheet,Daghero,Strijkers,Romeo1,Romeo2} reporting unusual conductance features, namely conductance dips and anomalous values of the zero-bias conductance (ZBC), not expected in the standard BTK model.

From a mathematical point of view, including a Dirac delta potential within the BdG formalism, as done in the BTK approach, is equivalent to impose matching conditions for the scattering wavefunctions diagonal in the particle-hole representation\cite{deGennes}. However, off-diagonal boundary conditions are also mathematically allowed by the BdG formalism and they could account for the appearance of anomalous features in the conductance spectra for N/S junctions.

In this paper we extend the BTK approach to include non-diagonal boundary conditions in the BdG scattering problem, introducing an interface potential that mixes electron and hole components of the BdG state. We show that this potential describes the proximity effect at the interface and it is responsible for the formation of conductance dips in the differential conductance spectra of N/S contacts. The comparison of the differential conductance curves with the experimental data suggests that a phase $\pi$-shift between the bulk and the interface order parameter, probably due to the presence of a localized magnetic moment at the interface, is at the origin of the conductance dips. The latter hypothesis is carefully analyzed by using a discretized model of the N/S junction.

The paper is organized as follows: in Sec. II we formulate the continuous model of the N/S interface introducing a particle-hole mixing term in the scattering potential. Off-diagonal boundary conditions are derived and the scattering coefficients are analytically determined. In Sec. III we show the differential conductance curves for N/S contacts obtained using generalized boundary conditions. We compare temperature evolution of conductance spectra with the existing theoretical models and available experimental data. Possible phase shift effects at the interface are discussed. In Sec. IV we use a discretized model to analyze the phase shift formation and evaluate the necessary physical conditions to observe it. Conclusions are given in Sec. V.

\section{Model}
We consider a one-dimensional N/S junction described by the BdG equations
\begin{eqnarray}
\left[\mathcal{H} + V\left(x\right)\right]\psi\left(x\right) = E \psi\left(x\right),
\end{eqnarray}
which completely define the quasi-particle state $\psi\left(x\right) = (u_{\uparrow}\left(x\right), u_{\downarrow}\left(x\right), v_{\uparrow}\left(x\right),v_{\downarrow}\left(x\right))^{t}$ having excitation energy $E$ above the Fermi energy $E_{F}$.
The Hamiltonian $\mathcal{H}$, which describes the bulk properties of the junction is
\begin{equation}
\label{eq.1}
\mathcal{H}=\left(\begin{array}{cc}
\hat{H}_{0} & \Delta(x)\emph{i} \hat{\sigma}_{y} \\
-\Delta^{*}(x)\emph{i}\hat{\sigma}_{y} & -\hat{H}^{*}_{0}
\end{array} \right),
\end{equation}
with
\begin{equation}
\hat{H}_{0} =\left[-\dfrac{\hbar^{2}\partial^{2}_{x}}{2m}-E_{F}\right]\mathbb{\hat{I}},
\end{equation}
where $\mathbb{\hat{I}}$ represents the identity operator in the spin space and $\hat{\sigma}_{y}$ is the Pauli matrix. We assume that the Fermi energy $E_{F}$ and the effective mass $m$ in the normal side of the junction ($x<0$) are equal to those in the superconductor ($x>0$), while the superconducting order parameter is taken of the form $\Delta(x)= \Delta \theta(x)$, where $\theta(x)$ is the Heaviside step function. Differently from the standard BTK treatment, we model the potential barrier at the interface ($x=0$) by a particle-hole mixing operator
\begin{equation}
\label{eq.5}
V\left(x\right)=\left(\begin{array}{cc}
U_{0}\mathbb{\hat{I}} & \emph{i} U_{1}  \hat{\sigma}_{y} \emph{e}^{\emph{i}\varphi} \\
-\emph{i} U_{1}  \hat{\sigma}_{y} \emph{e}^{-\emph{i}\varphi} & - U_{0}\mathbb{\hat{I}}
\end{array} \right)\delta \left(x\right),
\end{equation}
where $U_{0}$ indicates the usual BTK barrier strength, while the term $U_{1}$ describes the interfacial electron-hole coupling strength. The off-diagonal components of $V(x)$ describe the presence of a weak superconducting interface\cite{Tinkham} of negligible thickness compared to the coherence length of the superconductor. The variable $\varphi$ represents the phase difference between the interface and the bulk superconducting order parameter. Maintaining arbitrary values of $\varphi$, a Josephson current $I_{J}(\varphi) \propto \sin(\varphi)$\cite{Barone} is expected to flow through the interface. The free energy of the system is expected to be minimized when Josephson current vanishes, i.e. for $\varphi=0$ or $\pi$, the value $\varphi=0$ being a free energy minimum of the N/S junction. On the other hand, the value $\varphi=\pi$ can become an energy minimum if a magnetic moment is formed at the interface (e.g., transition metals easily oxidize producing localized magnetic states). Indeed, in the presence of a sufficiently strong magnetic moment, the interfacial phase can be modified from $0$ to $\pi$, the sign change of the interfacial order parameter following a mechanism similar to the one described in Ref. [\onlinecite{Kontos}].

In the following, we calculate the differential conductance of the N/S junction by considering the generalized boundary conditions of the scattering problem. The wave function of an electron with spin $\sigma =\{\uparrow, \downarrow\}$ coming from the N-side of the junction is given by:
\begin{eqnarray}
\label{eq.2}
\psi_{N}^{\sigma}(x)  & = & \left(\begin{array}{c}
\delta_{\uparrow \sigma} \\
\delta_{\downarrow \sigma} \\
0 \\
0
\end{array}\right)\emph{e}^{\emph{i}kx} + r_{e}^{\uparrow}\left(\begin{array}{c}
1 \\
0 \\
0 \\
0
\end{array}\right)\emph{e}^{-\emph{i}kx} \nonumber \\
& + & r_{e}^{\downarrow}\left(\begin{array}{c}
0 \\
1 \\
0 \\
0
\end{array}\right)\emph{e}^{-\emph{i}kx} + r_{h}^{\uparrow}\left(\begin{array}{c}
0 \\
0 \\
1 \\
0
\end{array}\right)\emph{e}^{\emph{i}qx} \nonumber \\
& + & r_{h}^{\downarrow}\left(\begin{array}{c}
0 \\
0 \\
0 \\
1
\end{array}\right)\emph{e}^{\emph{i}qx}.
\end{eqnarray}
Here the coefficients $r_{e}^{\uparrow,\downarrow}$ and $r_{h}^{\uparrow,\downarrow}$ correspond, respectively, to normal reflection and Andreev reflection, while $\hbar k = \sqrt{2m\left(E_{F}+E\right)}$ and $\hbar q = \sqrt{2m\left(E_{F}-E\right)}$ indicate the electron and hole wave vectors.\\
 In the superconducting region, we have
\begin{eqnarray}
\label{eq.3}
\psi_{S}(x)  & = & t_{e}^{\uparrow}\left(\begin{array}{c}
u \\
0 \\
0 \\
v
\end{array}\right)\emph{e}^{\emph{i}k_{+}x}
 +  t_{e}^{\downarrow}\left(\begin{array}{c}
0 \\
u \\
-v \\
0
\end{array}\right)\emph{e}^{\emph{i}k_{+}x} \nonumber \\ &+& t_{h}^{\downarrow}\left(\begin{array}{c}
v \\
0 \\
0 \\
u
\end{array}\right)\emph{e}^{-\emph{i}k_{-}x}
 +  t_{h}^{\uparrow}\left(\begin{array}{c}
0 \\
v \\
-u \\
0
\end{array}\right)\emph{e}^{-\emph{i}k_{-}x},
\end{eqnarray}
where the coefficients $t_{e}^{\uparrow}$, $t_{e}^{\downarrow}$, $t_{h}^{\uparrow}$, $t_{h}^{\downarrow}$ correspond to the transmission as electron-like  and hole-like quasiparticle with wave vectors $\hbar k_{\pm} = \sqrt{2 m\left(E_{F} \pm \sqrt{E^{2} - \Delta^{2}}\right)}$, the BCS\cite{BCS} coherence factors being
\begin{equation}
\label{eq.4}
u^{2} = 1 - v^{2} = \frac{1}{2}\left(1 + \frac{\sqrt{E^{2} - \Delta^{2}}}{E}\right).
\end{equation}
The coefficients in Eqs. (\ref{eq.2}) and (\ref{eq.3}) can be determined by using the generalized boundary conditions for the wave functions at the interface:
\begin{eqnarray}
\label{eq.matching}
\psi_{N}^{\sigma}(0) & = & \psi_{S}(0) \\ \nonumber
\partial_{x}\psi_{S}\vert_{x=0} & - & \partial_{x}\psi_{N}^{\sigma}\vert_{x=0}=\, \mathcal{A}\,\psi_{S}(0).
\end{eqnarray}
The matching matrix
\begin{equation}
\label{eq.matching-matrix}
\mathcal{A}= k_{F}Z_{0} \mathbb{\hat{I}}_{4 \times 4}+ k_{F} Z_{1}\left(\begin{array}{cc}
0 & \emph{i}\hat{\sigma}_{y}\emph{e}^{\emph{i}\varphi}\\
\emph{i}\hat{\sigma}_{y}\emph{e}^{-\emph{i}\varphi} & 0 \\
\end{array}\right)
\end{equation}
contains a diagonal term in the particle-hole representation with the usual BTK parameter $Z_{0}=2mU_{0}/(\hbar^{2}k_{F})$, and off-diagonal terms of strength $Z_{1}=2mU_{1}/(\hbar^{2}k_{F})$. The Eqs. (\ref{eq.matching})-(\ref{eq.matching-matrix}) provide the simplest particle-hole mixing boundary conditions mathematically allowed by the BdG formulation.\\
Using the above boundary conditions on the wave functions and the Andreev approximation $(k_{+} = k_{-} = k = q = k_{F})$, we find
the following expression for the scattering coefficients assuming the injection of a spin-up electron from the normal side (the result doesn't depend on the spin of the incoming process)
\begin{eqnarray}
\label{eq.6}
r_{h}^{\downarrow} & = & \dfrac{4uv-2\emph{i}\emph{e}^{-\emph{i}\varphi}Z_{1}(u^{2}-v^{2})}{4u^{2}+4\emph{i}uvZ_{1}\cos\varphi+(u^{2}-v^{2})(Z_{0}^{2}+Z_{1}^{2})} \\ \nonumber
r_{e}^{\uparrow} & = & -\dfrac{4\emph{i}uvZ_{1}\cos \varphi+(u^{2}-v^{2})[Z_{0}(2\emph{i}+Z_{0})+Z_{1}^{2})]}{4u^{2}+4\emph{i}uvZ_{1}\cos\varphi+(u^{2}-v^{2})(Z_{0}^{2}+Z_{1}^{2})} \\ \nonumber
t_{e}^{\uparrow} & = & \dfrac{4u-2\emph{i}(uZ_{0}-vZ_{1}\emph{e}^{-\emph{i}\varphi})}{4u^{2}+4\emph{i}uvZ_{1}\cos\varphi+(u^{2}-v^{2})(Z_{0}^{2}+Z_{1}^{2})} \\ \nonumber
t_{h}^{\downarrow} & = & \dfrac{2\emph{i}(vZ_{0}-uZ_{1}\emph{e}^{-\emph{i}\varphi})}{4u^{2}+4\emph{i}uvZ_{1}\cos\varphi+(u^{2}-v^{2})(Z_{0}^{2}+Z_{1}^{2})},
\end{eqnarray}
while the absence of spin-flip scattering implies $r_{e}^{\downarrow} =r_{h}^{\uparrow}=t_{e}^{\downarrow}=t_{h}^{\uparrow}=0$. Once the scattering coefficients are obtained, we can calculate the differential conductance by the formula \cite{BTK}
\begin{eqnarray}
\label{eq.7}
G(V) \propto \sum_{\sigma}\int dE \left[1+A_{\bar{\sigma}} -B_\sigma \right]\left[-\dfrac{\partial f(E-\emph{e}V)}{\partial E}\right]
\end{eqnarray}
where  $A_{\sigma} = \vert r_{h}^{\sigma}\vert^{2}$ and $B_{\sigma} = \vert r_{e}^{\sigma}\vert^{2}$ are the Andreev reflection and normal reflection probabilities, respectively, $f(E)$ is the Fermi-Dirac distribution, while the notation $\bar{\sigma}$ indicates the spin polarization opposite to $\sigma$.

\section{Results}

We first study the finite temperature conductance spectra of the N/S junction emphasizing the effects of the barrier strengths $Z_{0}$, $Z_{1}$ and of the phase $\varphi$.
\begin{figure}[htbp]
\centering
\includegraphics[scale=0.366,clip=]{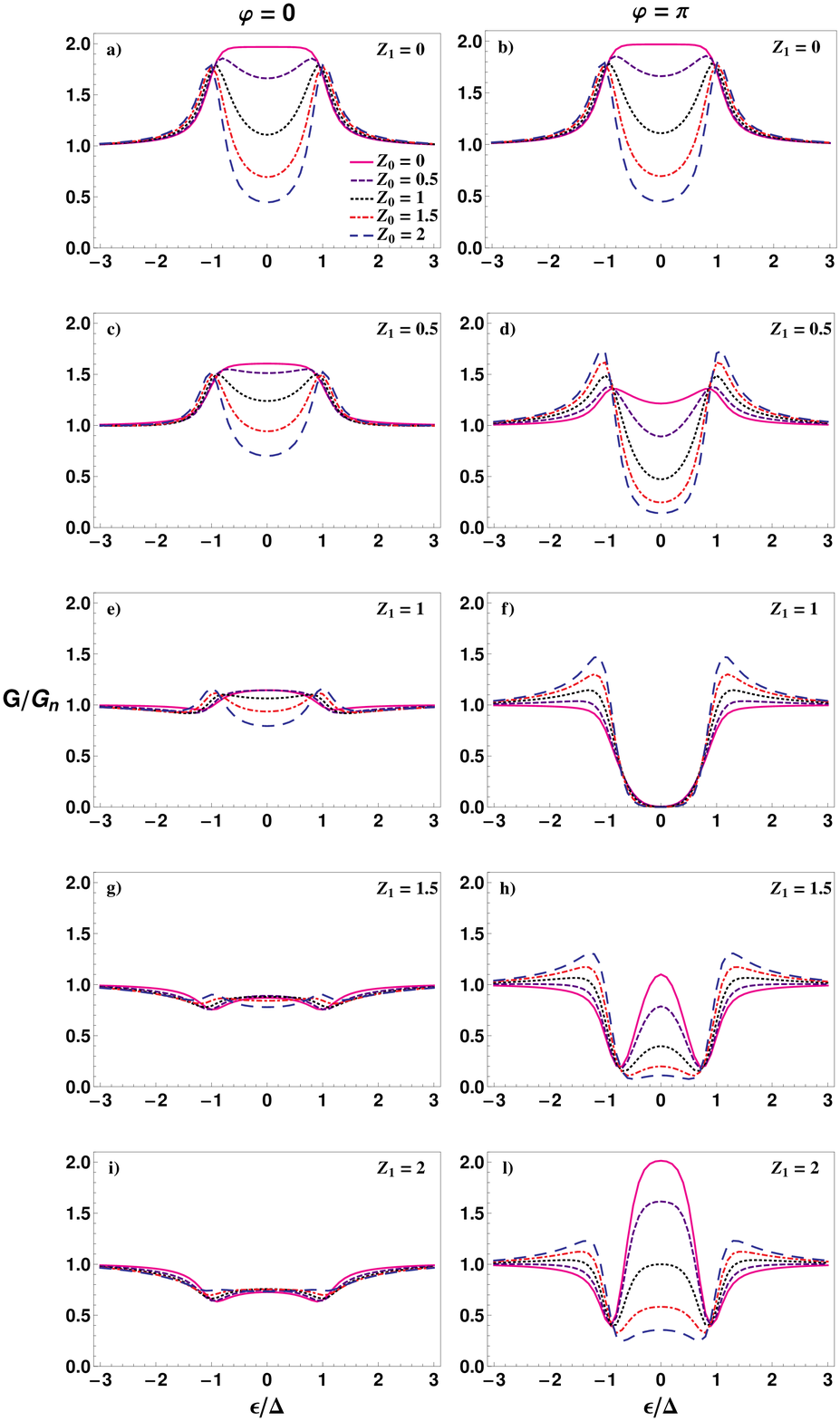}
\caption{(Color online) Normalized differential conductance curves, $G/G_{n}$ \textit{vs} $\epsilon/\Delta$, calculated at $T = 1.6 K$ from Equation (11). The different curves are obtained for distinct values of $Z_{0}$, $Z_1$ and $\varphi$ (values in the panels).}
\label{fig1}
\end{figure}
In Figure \ref{fig1} we show the normalized conductance $G/G_{n}$ \textit{vs} $\epsilon/\Delta$, with $G_{n} = G(eV\gg\Delta)$, for different values of $Z_0$ and $Z_1$ at a fixed temperature $T=1.6 K$, computed by using Eq. (\ref{eq.7}). Two cases are considered: (i) $\varphi=0$, shown in the left panels; (ii) $\varphi=\pi$, shown in the right panels. In each plot, different curves correspond to different $Z_0$ values (ranging from 0 to 2), while $Z_1$ is fixed as labelled. \\
For $\varphi=0$ and $Z_{1}=0$ (Figure 1(a)) the usual BTK behavior is recovered. In this case, the zero-bias conductance is suppressed as $Z_{0}$ is increased, while two peaks at $\epsilon/\Delta \approx \pm 1$ appear. Fixing $Z_{1}=0.5$ (Fig. 1(c)), we observe a reduction of the amplitude of the zero bias conductance feature compared to the $Z_{1}=0$ case. The difference between the conductance lowering induced by $Z_{0}$ and the peculiar amplitude reduction induced by $Z_{1}$ is evident: while the increasing of $Z_{0}$ induces a zero-bias conductance minimum, a tendency to increase the zero-bias conductance is observed by rising $Z_{1}$ to 1.0 (Fig. 1(e)), $Z_{1}=1.5$ (Fig. 1(g)) and $Z_{1}=2$ (Fig. 1(i)).\\
A different scenario is observed for $\varphi=\pi$: the effect of moderate values of $Z_{1}$, namely $Z_{1}=0.5$ (Fig. 1(d)) and $Z_{1}= 1$ (Fig. 1(f)), combine with $Z_{0}$ to give a relevant effective barrier strength leading to a strong suppression of the sub-gap conductance up to fully gapped spectra. For $Z_{1}=1.5$ (Fig. 1(h)) and $Z_{1}=2$ (Fig. 1(l)) an evident zero-bias peak  surrounded by two dips at $\epsilon/\Delta \approx \pm 1$ appears. Such ZBC peak exists for all $Z_{0}$ values in the range $[0,2]$, the junction transparency reduction having effect only on the peak amplitude.\\
All the conductance structures presented above (coming from the generalized boundary conditions) cannot be recovered within the standard BTK approach (except for the case $Z_1$=0). Moreover, the interface potential given in Equation (\ref{eq.5}) can be further generalized to include, spin-orbit interaction in the plane perpendicular to the transport direction, local magnetic moments and triplet or non-centrosymmetric  superconducting correlations. The above complications make the interface potential an off-diagonal differential operator of the form $\mathcal{B}(x,\partial_{x,y,z})\delta(x)$ acting on the Nambu space which induces an extended class of particle-hole mixing boundary conditions. Extending the BTK theory along this direction produces analytic results for the scattering coefficients which can be directly employed to explain anomalous conductance spectra.\\
In Figure 2 we compare the temperature evolution of the conductance spectra as obtained for $\varphi = \pi$ in the generalized BTK model introduced above, with the two-gap model \cite{Strijkers}. In particular, in Fig. 2(a) we show theoretical curves calculated in the temperature range between 0.1 K and 5.1 K by assuming $Z_0$=0.35, $Z_1$=2, $\Delta_{Nb}$=1.5 meV, $\varphi=\pi$, while in Fig. 2(b) the conductance curves are obtained by considering the parameters $Z$=0.14, $\Delta_1$=0.99 meV, $\Delta_2$=0.47 meV, in the two-gap model. Both models can be used to reproduce (black solid lines in Fig. 2(a) and 2(b)) the experimental data reported for $Cu/Nb$ contacts in Ref. [\onlinecite{Strijkers}]. For both the theoretical curves is necessary to fix a temperature value of $0.9 K$, that is lower than the bath temperature and it has been motivated as the result of non-equilibrium effects \cite{Strijkers} as well as in terms of other physical effects \cite{Romeo2}.
However, the temperature evolution of the conductance spectra shows appearance of different low temperature features at zero bias as well as at the gap edge. Moreover, the temperature dependence of the structures, namely ZBC and $E_{dip}$, evolve differently. By rising the temperature, a non-monotonic evolution of $E_{dip}$ (Fig. 2(c))  and a faster reduction of ZBC (Fig. 2(d)) is observed for the case of Fig. 2(b). As a consequence, very low temperature experiments are necessary to distinguish the two models, and thus understanding the physical origin of the anomalous conductance features observed in the point contact experiments in N/S devices.
It is worth to notice that both scenarios recalled by the two models are physically plausible. Indeed, Nb and Cu oxides are known to exhibit magnetic correlations that could realize effective local polarization enabling a phase shift at the interface (discussed below); on the other hand, formation of a (proximized) weak superconducting layer at the N/S interface is always possible.

\begin{figure}[!t]
\centering
\includegraphics[scale=0.8,clip=]{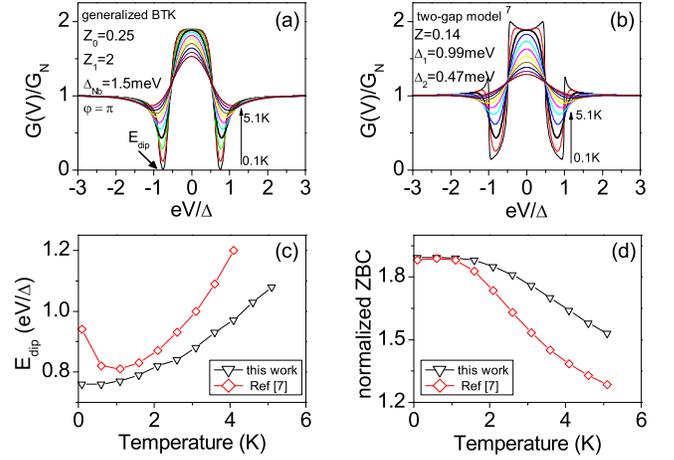}
\caption{Temperature evolution of the normalized conductance curves obtained for (a) generalized BTK model with particle-hole mixing boundary conditions and for (b) two-gap model\cite{Strijkers} for the parameters $Z$=0.14, $\Delta_1$=0.99 meV (bulk gap), $\Delta_2$=0.47 meV (proximized gap). The thick black line in the two plots represent the best fit for $Cu/Nb$ experimental data reported in Ref. [\onlinecite{Strijkers}]. (c) Comparison of the temperature evolution of the energy position of the conductance dips, $E_{dip}$ vs T, obtained from (a) and (b). (d) Comparison of the temperature evolution of the ZBC, obtained from (a) and (b).}
\label{fig2}
\end{figure}

Up to now we have assumed that the interface phase $\varphi=\pi$ is the result of an emergent magnetic moment at the interface even though the Hamiltonian model (Eq. (3)) does not include explicitly magnetic correlations. In the following, we prove that also modifying the interface potential $V\left(x\right)$ (Eq. (\ref{eq.5})) by adding a localized magnetic potential of the form $U_{2} \hat{\sigma}_{z}\otimes\hat{\sigma}_{z} \delta(x)$, with dimensionless strength $Z_{2}$ does not change the peculiar shape of the conductance given in Fig. \ref{fig2}.
\begin{figure}[!h]
\centering
\includegraphics[scale=0.63,clip=]{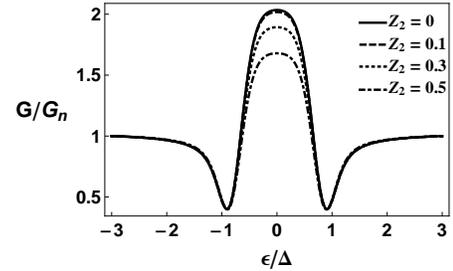}
\caption{Normalized differential conductance curves, $G/G_{n}$ \textit{vs} $\epsilon/\Delta$ calculated by considering the inclusion of a localized magnetic contribution to the potential $V\left(x\right)$. Parameters are $T = 1.6 K$ for $Z_{0}=0$, $Z_{1}=2$ and $\varphi =\pi$, with $Z_{2} \in [0,0.5]$.}
\label{fig3}
\end{figure}
Assuming an incoming spin-up electron, the expressions of the Andreev and normal reflection become:
\begin{eqnarray}
r_{h}^{\downarrow} & = & \dfrac{4uv-2\emph{i}\emph{e}^{-\emph{i}\varphi}Z_{1}(u^{2}-v^{2})}{4u^{2}+4\emph{i}uvZ_{1}\cos\varphi+(u^{2}-v^{2})Z^{2}_{+}} \\ \nonumber
r_{e}^{\uparrow} & = & -\dfrac{4\emph{i}uvZ_{1}\cos \varphi+(u^{2}-v^{2})[Z^{2}_{+} + 2 \emph{i} (Z_{0} + Z_{2})]}{4u^{2}+4\emph{i}uvZ_{1}\cos\varphi+(u^{2}-v^{2})Z^{2}_{+}},
\end{eqnarray}
where $Z^{2}_{+} = Z_{1}^{2} + (Z_{0}+Z_{2})^{2} $. Analogous expressions, characterized by the parameter $Z^{2}_{-}= Z_{1}^{2} + (Z_{0}-Z_{2})^{2}$, are found for $r^{\uparrow}_{h}$ and $r^{\downarrow}_{e}$ in the case of the scattering problem of a spin-down electron coming from the normal side. It is also worth mentioning that the magnetic moment makes different the Andreev (and the normal) reflection coefficients of scattering processes originated by electron-like quasiparticles of opposite spin projection (i.e. $r^{\downarrow}_{e/h} \neq r^{\uparrow}_{e/h}$), while spin-flipping reflection processes are not allowed by the Hamiltonian structure.
In Fig. \ref{fig3} the effect of $Z_{2}$ on the conductance curves is shown. Increasing $Z_{2}$, the amplitude of the zero-bias peak is lowered, while the conductance features are almost the same for $Z_{2} \in [0,0.5]$ except for a small reduction of the ZBC peak amplitude. This shows that adding a localized magnetic term does not change the conductance shape. The latter property derives from the fact that the magnetic correlation only renormalizes the interface potential ($Z_{0}\rightarrow Z_{0}\pm Z_{2} $) producing a spin-sensitive effective barrier, whose effects are difficult to be distinguished from the ones expected in non-magnetic case. This explains why the presence of an emergent magnetic moment at the interface is difficult to be experimentally confirmed by means of point contact spectroscopy.

\section{Discretized model}
As we have seen in Figs. 1 and 2 the conductance dips appear in the generalized BTK approach for  $\varphi=\pi$ and this phase value can be associated to a localized magnetic moment. In fact, the presence of a localized magnetic moment at the interface can make a phase gradient of $\pi$ energetically favorable. In order to identify the physical conditions (interface polarization and transparency) to realize the $\pi$-shift, we consider a discretized formulation that allows to describe spatial dependent potentials without increasing the computational complexity.
We model a system with an odd number of sites $N$ in which $(N-1)/2$ sites are used for both the normal and the superconducting side, while one normal site with magnetic ($\Gamma$) and non-magnetic ($U$) potentials is assigned to the interface (see Fig. 4). The nearest-neighbor hopping parameter $t = \hbar^{2} / (2 m a^{2})$, expressed in terms of the sites distance $a$, is assumed homogeneous and it used as energy unit, fixing $t \simeq 10 \cdot\Delta_{Nb}$ in order to have $\xi_{Nb} \simeq 10\cdot a$. Temperature is measured in dimensionless units $\tau=k_{B}T/t$. Hereafter, we set $t= 16.2$ meV as the energy cut-off of the theory, the latter being of the same order of magnitude of the Debye energy $\hbar \Omega_D$. This choice guarantees that only states with phonon-mediated attraction (i.e. with $\epsilon \in [0, \hbar \Omega_{D}]$) are retained, simultaneously ensuring the long wavelength limit of the considered wave functions. Under these assumptions, the relevant wave functions present a De Broglie wavelength greater than the lattice constant $a$, while the associated eigenvalues defines a near-parabolic energy dispersion.
\begin{figure}[!t]
\centering
\includegraphics[scale=0.89,clip=]{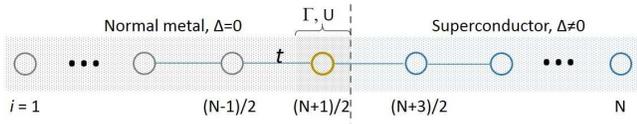}
\caption{Discretized model of the N/S junction consisting of $N$ (odd) sites: (N+1)/2 normal sites ($\Delta =0$) and (N-1)/2 superconducting sites ($\Delta \neq 0$); magnetic ($\Gamma$) e non-magnetic ($U$) potentials are present at the interface site $i=(N+1)/2$. The hopping parameter $t$ is homogeneous along the system.}
\label{fig4}
\end{figure}

\begin{figure}[!t]
\centering
\includegraphics[scale=0.8,clip=]{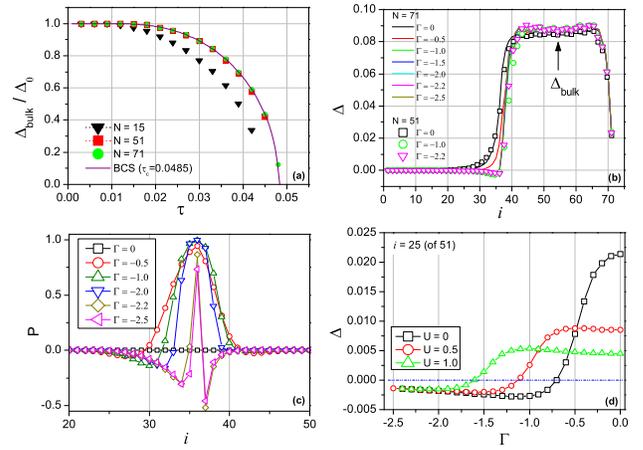}
\caption{Numerical results obtained in the discretized model. (a) Temperature dependence of the bulk superconducting gap for different systems, namely $N=15$, $N=51$, $N=71$, assuming transparent barrier, $U=0$. Numerical data, normalized to the low temperature value $\Delta_0$, are compared to theoretical behaviour expected in the BCS model for $\tau_c = 0.0485$. (b) Spatial dependence of the superconducting gap for 71-sites and 51-sites systems, by assuming $U=0$, a constant BCS coupling, $\lambda_i = \lambda$,  and dimensionless temperature $\tau =0.025$. Different curves correspond to different values of $\Gamma$. Lines refer to data obtained for 71-sites system; scattered symbols refer to 51-sites system, rescaled to compare the data sets. The arrow indicates the region where the superconducting gap is calculated self-consistently. (c) Spatial dependence of the polarization calculated for $N=71$ for different $\Gamma$ values, with $U=0$ and $\tau =0.025$. (d) Effect of the barrier strength $U$: the superconducting gap vs $\Gamma$ is evaluated at the site $i=25$ (of 51), at $\tau =0.025$ for three different transparency conditions ($U=0$, $U=0.5$, $U=1$), the magnetic site being located at $i=26$.}
\end{figure}

The discretized version of the BdG equations, in the presence of a Zeeman term $\Gamma(x)\hat{\sigma}_{z}$ added to the single-particle Hamiltonian $\hat{H}_{0}$ to account for the effective polarization at the interface, correspond to the following matrix equations ($\forall \ i \in[1,N]$):
\begin{eqnarray}
\label{eq.8}
\mathcal{M}^{(\sigma)}_{i}\Psi^{(\sigma)}_{i}+T\left(\Psi^{(\sigma)}_{i+1}+\Psi^{(\sigma)}_{i-1}\right)=0 ,
\end{eqnarray}
where ($\sigma=\pm$, $\pm \equiv \uparrow/\downarrow $)
\begin{eqnarray}
\mathcal{M}^{(\sigma)}_{i}&=&\left(\begin{array}{cc}
\varepsilon_{i}-E +\sigma \Gamma_{i}& \sigma \Delta_{i} \\
\sigma \Delta^{*}_{i} & -\varepsilon_{i}-E+\sigma \Gamma_{i}
\end{array}\right),
\end{eqnarray}
while $T=-t \  \hat{\sigma}_{z}$. Here $\varepsilon_i/t = 2  + U\delta_{i,(N+1)/2}$ is the energy of the $i$-th lattice site, while  $\Psi_i^{(\sigma)}=\left(u_{\sigma, i},v_{\bar{\sigma},i}\right)^{t}$  is the discretized BdG state in the absence of spin-flip scattering. $\Gamma_{i}/t = \Gamma \delta_{i,(N+1)/2}$ is the site dependent Zeeman energy that we take different from zero only at the interface site.
Using Dirichlet boundary conditions $\Psi^{(\sigma)}_{1}=\Psi^{(\sigma)}_{N}=0$, we get electron-like eigenstates
\begin{equation}
\Phi^{(\sigma , n)}= \sum_{i=1}^N \mathcal{A}_i \otimes  \left(u_{\sigma, i}^{(n)},v_{\bar{\sigma},i}^{(n)}\right)^{t}
\end{equation}
associated to positive energy eigenvalues ($\epsilon_n \geq 0$), with $\mathcal{A}_i=(\delta_{1,i},...,\delta_{N,i})^t$. The spatial dependence of the superconducting gap is computed as\cite{nota}
\begin{equation}
\Delta_i =\frac{\lambda_i}{2} \sum_{n} \left[ u_{\uparrow ,i}^{(n)}  v^{(n)\star}_{\downarrow ,i}-u_{\downarrow ,i}^{(n)}  v^{(n)\star}_{\uparrow ,i}\right] \tanh \left( \frac{\epsilon_n}{2 k_BT}\right),
\label{eq15}
\end{equation}
the sum being calculated for $\epsilon_n \in [0,\hbar\Omega_D]$. The attractive phonon-mediated local potential $\lambda_i$ is assumed constant\cite{lambda}  ($\lambda_i= \lambda$) also in the normal side of the junction to take into account for the proximity effect and the interdiffusion of atoms belonging to the N- and S-side.
We consider the bulk superconducting gap  $\Delta_{bulk}$ at the center of the S-region, in order to avoid finite size effects. $\Delta_{bulk}$ is self-consistently computed using Eq. (\ref{eq15}) with accuracy better than $1\%$ starting from $\Delta_i = 0.15 t$ in S, and by fixing $\lambda = 2.4 \ t$. The polarizing effect of the magnetic site at the interface can be quantified defining the site-dependent polarization $P_i= (n_{i,\uparrow}-n_{i,\downarrow})/(n_{i,\uparrow}+n_{i,\downarrow})$, where
\begin{eqnarray}
n_{i,\uparrow} + n_{i,\downarrow} & = & \sum_n \left[ |u_{\uparrow ,i}^{(n)}|^2 f_n + |v_{\downarrow ,i}^{(n)}|^2 (1-f_n) \right]  +\\ \nonumber
& + & \sum_n \left[ |u_{\downarrow ,i}^{(n)}|^2 f_n + |v_{\uparrow ,i}^{(n)}|^2 (1-f_n) \right]\\  \nonumber
\end{eqnarray}
and
\begin{eqnarray}
n_{i,\uparrow} - n_{i,\downarrow} & = & \sum_n \left[ |u_{\uparrow ,i}^{(n)}|^2 f_n - |v_{\downarrow ,i}^{(n)}|^2 (1-f_n) \right]  +\\ \nonumber
& + & \sum_n \left[ -|u_{\downarrow ,i}^{(n)}|^2 f_n + |v_{\uparrow ,i}^{(n)}|^2 (1-f_n) \right],\\ \nonumber
\end{eqnarray}
with $f_n = f(\epsilon_n)$ a shortened notation standing for the Fermi-Dirac distribution. In order to capture the bulk-like behaviour using a finite size system, the system size $N$ has been progressively increased from $N=15$ to $N=71$, while monitoring the temperature dependence of $\Delta_{bulk}$.
The results of this analysis are shown in Figure 5(a) where normalized values of $\Delta_{bulk}$ are presented as a function of the dimensionless temperature $\tau$. For a system size of $N=15$ a size-induced suppression of the superconducting gap is observed, this effect being more evident close to the transition temperature $\tau_c$. Increasing the system size up to $N=51$ produces a $\Delta_{bulk}$ vs $\tau$ curve very close to the one obtained for the $N=71$ case, signaling that the bulk limit of the interface model has been reached. The temperature evolution of $\Delta_{bulk}$ for $N=71$ has been compared with the BCS behaviour giving a dimensionless critical temperature $\tau_c=0.0485$, corresponding to a niobium critical temperature $T_c^{Nb}\simeq 9.1 K$. In Figure 5(b) we present the spatial dependence of the superconducting gap (for system size $N=51$ and $N=71$) fixing the Zeeman energy $\Gamma$ of the magnetic potential in the range $[-2.5,0]$, while taking $U=0$ (transparent interface) and $\tau=0.025$ (i.e. $T \sim T^{Nb}_c/2$). In order to compare spectra obtained for systems with different size, the data referring to $N=51$ have been rescaled.
For $-1.0<\Gamma <0$ we observe ordinary proximity effect where finite superconducting order parameter is induced in the N-side on a length of about $10 a \simeq \xi_{Nb}$. For sufficiently strong magnetization $\Gamma \leq -1.0$ negative order parameter is induced on the same length scale.
Reduction of the order parameter on the right border is due to the  S/vacuum interface.
In Figure 5(c) we show the spatial dependence of the polarization, calculated for $\Gamma \in [-2.5,0]$ and $U=0$. The polarizing effect of the localized magnetic moment asymmetrically extends on a distance of about $20 a$. In the superconducting side the induced polarization is inverted for large $\Gamma$ values\cite{bergeret05} ($\Gamma <-2.0$). The general aspect of the polarization curves evidences Friedel density oscillations.  We have also verified the effect of barrier strength on the inversion of the superconducting order parameter. In Figure 5(d), we show the gap value calculated (at site $i = 25$) in proximity of the interface, for a system size $N=51$, as a function of $\Gamma$, with enhanced resolution (step 0.1). For reduced transparency ($U>0$) a larger magnetic moment is necessary to induce the inversion of the interface order parameter.
The analysis of the pairing potential $\Delta_i$ shows that, in presence of a local polarization at the interface, a phase gradient $\varphi = \pi$ can be stabilized. For relatively transparent junctions (i.e. described by small values of $U$) the sign change of the interface order parameter can be obtained with moderate polarization strength, while strong polarization values are needed for opaque interface with higher values of $U$. Thus the probability to observe an hidden magnetic moment at the interface accompanied by a phase gradient is enhanced in transparent systems. The physical origin of a local magnetic moment at the $Cu/Nb$ interface probably resides in many-body effects which can be accounted for in the framework of the Anderson impurity model\cite{appelbaum}.

\section{Conclusions}
We have generalized the BTK theory to include particle-hole mixing boundary conditions in the scattering problem, reporting analytic results for the scattering coefficients. We calculated the finite-temperature differential conductance spectra for N/S junctions showing the formation of conductance dips in the case of a phase $\pi$-shift at the interface. We demonstrated that the temperature evolution of the conductance spectra can discriminate the physical origin of the conductance dips, either the formation of a localized magnetic moment or the presence of a weak proximized superconducting layer at the interface. According to the analysis, a localized magnetic moment could make a sign change of the superconducting order parameter energetically favorable. Finally, we used a discretized model to determine the necessary physical conditions under which the $\pi$-shift is realized: transparent interfaces can easily sustain a phase gradient as the effect of a weak interface magnetization, while for reduced transparencies a relative strong localized magnetization would be necessary.

\begin{acknowledgments}
We thank A. Braggio and F. Giazotto for helpful discussions.
\end{acknowledgments}


\begin{thebibliography}{99}
\bibitem{nota} Notice that the spinorial particle-hole notation adopted in this work follows from the Bogoliubov field transformation $\psi_{\sigma}(x)=\sum_n [u_{n\sigma}(x) \gamma_n+v_{n\sigma}^{\ast}(x)\gamma_n^{\dagger}]$, with $x/a=i$. Using the Bogolons rapresentation the BCS Hamiltonian takes the form $H_{BCS}=\sum_n E_n \gamma^{\dagger}_n\gamma_n+E_g$.
\bibitem{lambda} The numerical results do not change taking $\lambda_i=\lambda \neq 0$ only in the S-region and in the proximized N-region whose extention is about $\xi \sim 10 \cdot a$.
\end{thebibliography}
\end{document}